\documentclass[a4paper,11pt]{article}
\usepackage{graphicx,amssymb,amstext,amsmath,mathabx,mathtools,esvect,xparse}
\usepackage{color,wrapfig}
\usepackage{floatflt}
\usepackage{lipsum}

\definecolor{cbl}{rgb}{0,0,1}               

\newcommand{\bc}{\begin{center}}
\newcommand{\ec}{\end{center}}
\def\ba#1{\begin{array}{#1}\displaystyle}
\newcommand{\ea}{\end{array}}

\newcommand{\beq}{\begin{equation}}
\newcommand{\eeq}{\end{equation}}
\newcommand{\beqa}{\begin{eqnarray}}
\newcommand{\eeqa}{\end{eqnarray}}

\newcommand{\bi}{\begin{itemize}}
\newcommand{\ei}{\end{itemize}}

\newcommand{\bra}{\langle}
\newcommand{\ket}{\rangle}

\newcommand{\TTb}{\mathrm{T}\overline{\mathrm{T}}}
\newcommand{\bal}{\boldsymbol{\alpha}}
\newcommand{\bol}{\boldsymbol{0}}
\newcommand{\bel}{\boldsymbol{\beta}}

\usepackage{cite}

\definecolor{purple_nice}{rgb}{0.4,0.2,0.7}
\definecolor{fuel_blue}{RGB}{42,162,185}
\definecolor{YInMn_blue}{RGB}{46, 80, 144}
\definecolor{ultramarine}{RGB}{63, 0, 255}
\definecolor{KLEIN_blue}{rgb}{0, 0.18, 0.65}
\usepackage[linktocpage=true]{hyperref}
\hypersetup{
    colorlinks=true,
    linkcolor=YInMn_blue,
    citecolor=ultramarine,
    filecolor=fuel_blue,
    urlcolor=KLEIN_blue,
}

\makeatletter
\renewenvironment{abstract}{%
      \begin{center}%
        {\bfseries \normalsize\abstractname\vspace{\z@}}
      \end{center}%
      \quotation}
    {\endquotation}
\makeatother
\usepackage[normalem]{ulem}  

\begin{document}

\begin{titlepage}
\title{Completing the Bootstrap Program for $\TTb$-Deformed\\ Massive Integrable Quantum Field Theories}
\author{Olalla A. Castro-Alvaredo{\color{red} {$^\heartsuit$}}, Stefano Negro{\color{blue} {$^\clubsuit$}} and Fabio Sailis{ {$^\diamondsuit$}}\\[0.3cm]}
\date{\small {\color{red} {$^\heartsuit$}} $^\diamondsuit$  Department of Mathematics, City, University of London, 10 Northampton Square EC1V 0HB, UK\\
\medskip
{\color{blue} {$^\clubsuit$}}  {Department of Mathematics, University of York, York YO10 5DD, UK} \\
}
\maketitle
\begin{abstract} 
In recent years a considerable amount of attention has been devoted to the investigation of 2D quantum field theories perturbed by certain types of irrelevant operators. These are the composite field $\TTb$ -- constructed out of the components of the stress-energy tensor -- and its generalisations -- built from higher-spin conserved currents.
The effect of such perturbations on the infrared and ultraviolet properties of the theory has been extensively investigated.
 In the context of integrable quantum field theories, a fruitful perspective is that of factorised scattering theory. In fact, the above perturbations were shown to preserve integrability. The resulting deformed scattering matrices -- extensively analysed with the thermodynamic Bethe ansatz -- provide the first step in the development of a complete bootstrap program.
In this paper we present a systematic approach to computing matrix elements of operators in generalised $\TTb$-perturbed models, based on employing the standard form factor program. Our approach is very general and can be applied to all theories with diagonal scattering. We show that the deformed form factors, just as happens for the $S$-matrix, factorise into  the product of the undeformed ones and of a perturbation- and theory-dependent term.
From these solutions, correlation functions can be obtained and their asymptotic properties studied. Our results set the foundations of a new research program for massive integrable quantum field theory perturbed by irrelevant operators. 
\end{abstract}


\vfill

\noindent 
{\Large {\color{red} {$^\heartsuit$}}} o.castro-alvaredo@city.ac.uk\\
{\Large {\color{blue} {$^\clubsuit$}}} stefano.negro@york.ac.uk\\
{\Large {{$^\diamondsuit$}}} fabio.sailis@city.ac.uk\\

\hfill \today
\end{titlepage}

\section{Introduction}
Generalised $\TTb$ deformations of 2D quantum field theory (QFT) \cite{Smirnov:2016lqw, Cavaglia:2016oda, Dubovsky:2012wk, Caselle:2013dra} admit a variety of interesting interpretations. From coupling the original QFT to two-dimensional topological gravity \cite{Dubovsky:2017cnj} or to random geometry \cite{ Cardy:2018sdv} to a state-dependent change of coordinates \cite{Conti:2018tca}, among others.  For these reasons, $\TTb$ perturbations and their generalisations have been intensely studied in 2D quantum field theory \cite{Smirnov:2016lqw, Cavaglia:2016oda, Conti:2018jho, Conti:2018tca, Conti:2019dxg, Dubovsky:2023lza}, in the framework of the ODE/IM correspondence \cite{Aramini:2022wbn} (see \cite{Dorey:2019ngq} for a review), employing  thermodynamic Bethe ansatz (TBA) \cite{Hernandez-Chifflet:2019sua, Camilo:2021gro, Cordova:2021fnr, LeClair:2021opx, LeClair:2021wfd, Ahn:2022pia, Dubovsky:2012wk, Caselle:2013dra}, perturbed CFT \cite{Guica:2017lia, Cardy:2018sdv, Cardy:2019qao, Aharony:2018vux, Aharony:2018bad, Guica:2020uhm, Guica:2021pzy, Guica:2022gts}, string theory \cite{Baggio:2018gct, Dei:2018jyj, Chakraborty:2019mdf, Callebaut:2019omt}, holography \cite{McGough:2016lol, Giveon:2017nie, Gorbenko:2018oov, Kraus:2018xrn, Hartman:2018tkw, Guica:2019nzm, Jiang:2019tcq,Jafari:2019qns}, quantum gravity \cite{Dubovsky:2017cnj, Dubovsky:2018bmo, Tolley:2019nmm, Iliesiu:2020zld, Okumura:2020dzb, Ebert:2022ehb}, out-of-equilibrium CFT \cite{Medenjak:2020ppv,Medenjak:2020bpe}, in long-range spin chains \cite{Bargheer:2008jt,Bargheer:2009xy,PJG, Marchetto:2019yyt}, and in the context of the generalised hydrodynamics approach \cite{DDY,Cardy:2020olv,DHY}. A generalisation of this family of deformations has also been proposed for quantum-mechanical systems \cite{Gross:2019ach, Gross:2019uxi, ste_zeta} and higher-dimensional field theories \cite{Bonelli:2018kik, Taylor:2018xcy, Conti:2022egv}.

In this paper, we approach the 
 study of generalised $\TTb$ deformations from the viewpoint of integrable quantum field theory (IQFT) \cite{Zamolodchikov:1989hfa, Negro:2016yuu}, integrability techniques and the bootstrap program. Here we define the 
latter as consisting of the following steps: 1) computation of the exact scattering matrix, 2) computation of matrix elements (form factors) of local fields, 3) computation of correlation functions, and 4) consistency checks of those form factors and correlation functions, which include the TBA approach and Zamolodchikov's $c$-theorem \cite{Zamc}. To date  point 1) and a large part of point 4) (i.e. the TBA treatment) are very well developed, whereas points 2) and 3) and most of 4) are very much outstanding. In particular, although much work on the TBA analysis of the theories has been done,  it has not yet been possible to compare TBA results with information obtained from correlation functions.  In this paper, we show in great generality how these outstanding steps may be carried out for a large family of IQFTs, fields, and perturbations.

It is well known that in IQFT perturbations of $\TTb$-type lead  to a modification of the exact two-body scattering matrix by a multiplicative or CDD \cite{Castillejo:1955ed} factor.
The deformed $S$-matrix is
\begin{equation}
    S_{\boldsymbol{\alpha}}(\theta) = \Phi_{\boldsymbol{\alpha}}(\theta) S_{\boldsymbol{0}}(\theta)\;,
    \label{Smatrix}
\end{equation}
where $\boldsymbol{\alpha} = (\alpha_s)_{s\in\mathcal{S}\subset\mathbb{N}}$ is a, possibly infinite, vector, $S_{\boldsymbol{0}}(\theta)$ is some undeformed, consistent, factorised $S$-matrix in a theory with a single particle, and
\begin{equation}
    \Phi_{\boldsymbol{\alpha}}(\theta) = \exp\left[-i\sum_{s\in \mathcal{S}} \alpha_s \sinh(s\theta)\right]\;.
    \label{sum}
\end{equation}
A generalisation to more complicated spectra is possible along the same lines as presented here.
For simplicity, we have taken the fundamental mass scale $m=1$, but it is useful to notice that the combination $\alpha_s m^{2s}$ is dimensionless. $\mathcal{S}$ is a set of spin values, typically those of 
local conserved charges. Since $\Phi_{\boldsymbol{\alpha}}(\theta)$ is a CDD factor, the theory described by this new $S$-matrix is still integrable. 
The computation of the exact matrix elements (form factors) of local operators corresponding to the family of $S$-matrices (\ref{Smatrix})  has not  been attempted systematically. 
As we shall see, the unusual UV physics of $\TTb$-perturbed theories, will also emerge clearly when pursuing the form factor program. In this sense, form factors provide a new useful tool to understand this physics. For example, we will find parameter regions, where a distinct asymptotics of correlators can be read off from a form factor expansion and consistently interpreted in the context of the physics of $\TTb$ perturbations. Pursuing this program will not only shed light on the properties of the correlators of $\TTb$-perturbed IQFTs but also reveal the mathematical structure of a new family of solutions to the form factor equations. Form factors and correlation functions of a theory with similar features were first studied in \cite{sgMuss}. 
\vspace{0.1cm}

The main results of our work are: (1) We find closed solutions for the form factors of local and semi-local fields in generalised $\TTb$-perturbed IQFTs and, in some parameter regions, convergent two-point function expansions. (2) When all $\alpha_i<0$ two-point functions can be expanded in terms of form factors of the corresponding fields and the expansion is convergent both for long and short distances. For long distances (IR) the theory flows to a trivial (massive) fixed point, just as in the unperturbed case, whereas for short distance (UV) two-point functions still scale as power-laws and the $c$-theorem gives a finite result which is a function of the couplings $\boldsymbol{\alpha}$. (3) If $\alpha_s>0$, where $s$ is the largest spin involved in the sum (\ref{Smatrix}), the two-point function is divergent in the UV limit, and also in the IR limit for momenta above a certain threshold which is universally characterised by Lambert's $W$-function. 

\medskip
Our paper is organised as follows: In Section \ref{FFB} we review the form factor equations and minimal form factors in $\TTb$-perturbed theories. In Section \ref{factorisation} we present our main result, namely a derivation of the factorisation property of form factors, leading to close formulae for the form factors of $\TTb$-perturbed theories with diagonal scattering. In Section \ref{CFA} we discuss the asymptotics of correlation functions in different parameter regimes. In Section \ref{cTheorem} we discuss how Zamolodchikov's $c$-theorem can still evaluated for $\TTb$-perturbed theories, giving rise to a function that flows with the RG parameter $|\alpha|$. We conclude and discuss some open problems and subtleties of our derivation in Section \ref{conclu}. 

\section{Form Factor Bootstrap}
\label{FFB}
Let 
\beq  
F_n^{\mathcal{O}}(\theta_1,\ldots,\theta_n;\boldsymbol{\alpha})=\bra 0| \mathcal{O}(0)|\theta_1,\ldots, \theta_n\ket\,,
\eeq 
be a matrix element of a field $\mathcal{O}$ between the ground state $|0\ket$ and an $n$-particle in-state $|\theta_1,\ldots, \theta_n\ket$ characterised by rapidities $\{\theta_i\}_n$. Since we are considering only theories with a single particle we do not need to consider additional quantum numbers in this case. 
The form factor equations for a diagonal theory without bound states are \cite{KW,smirnovbook,mussardobook}
\beq 
F_n^{\mathcal{O}}(\theta_1,\ldots,\theta_i,\theta_{i+1},\ldots,\theta_n;\boldsymbol{\alpha})=S_{\bal}(\theta_{i}-\theta_{i+1}) F_n^{\mathcal{O}}(\theta_1,\ldots,\theta_{i+1},\theta_{i}, \ldots, \theta_n;\boldsymbol{\alpha})\,,
\label{FF1}
\eeq 
\beq 
F_n^{\mathcal{O}}(\theta_1+2\pi i, \theta_2, \ldots,\theta_n;\boldsymbol{\alpha})= \gamma_{\mathcal{O}}F_n^{\mathcal{O}}(\theta_2, \ldots, \theta_n,\theta_1;\boldsymbol{\alpha})\,,
\label{FF2}
\eeq 
 and 
\beq
 \lim_{\bar{\theta}\rightarrow \theta} (\bar{\theta}-\theta) F^{\mathcal{O}}_{n+2}(\bar{\theta}+i\pi, \theta, \theta_1,\ldots, \theta_n;\boldsymbol{\alpha}) = i (1-\gamma_{\mathcal{O}}\prod_{j=1}^n S_{\bal}(\theta-\theta_j)) F^{\mathcal{O}}_n(\theta_1,\ldots, \theta_n;\boldsymbol{\alpha})\,,
 \label{KRE}
 \eeq
with the two first equations constraining the monodromy of the form factors and the last equation, the kinematic residue equation, specifying their pole structure. In (\ref{KRE}) we find the parameter $\gamma_{\mathcal{O}}$ named the factor of local commutativity \cite{YZam,BB} which can be a non-trivial phase in theories possessing an internal symmetry (such as the Ising model which we have studied in more detail in \cite{longpaper}). The equations (\ref{FF1})-(\ref{KRE}) are typically solved by the following ansatz
\begin{equation}
    F_n^{\mathcal{O}}(\theta_1,\ldots,\theta_n;\boldsymbol{\alpha}) = H_{n}^{\boldsymbol{\alpha}} Q_n^{\mathcal{O}}(\theta_1,\ldots,\theta_n;\boldsymbol{\alpha}) \prod_{i<j}^n\frac{F_{\mathrm{min}}(\theta_{ij};\boldsymbol{\alpha})}{e^{\theta_i}+e^{\theta_j}}\;.
    \label{ansatz}
\end{equation}
where $\theta_{ij}:=\theta_i-\theta_j$ and $F_{\mathrm{min}}(\theta;\boldsymbol{\alpha})$ is the minimal form factor, that is, a minimal solution (i.e. without poles in the physical strip) of the form factor equations for two particles,
\beq
F_{\mathrm{min}}(\theta;\boldsymbol{\alpha})=S_{\boldsymbol{\alpha}}(\theta) F_{\mathrm{min}}(-\theta;\boldsymbol{\alpha})=F_{\mathrm{min}}(2\pi i-\theta;\boldsymbol{\alpha})\,.
\eeq 
Note that for spinless fields, form factors are functions of rapidity differences only, hence the minimal form factor depends on a single rapidity variable. For the same reason, one-particle form factors when non-vanishing, must be rapidity-independent. By construction, the product over minimal form factors in (\ref{ansatz}) provides a solution to equations (\ref{FF1}) and (\ref{FF2}). That means that the functions $Q_n^{\mathcal{O}}(\{\theta_i\}_n;\boldsymbol{\alpha})$ must be symmetric and $2\pi i$-periodic in all rapidities. In other words, they should be functions of the exponentials 
$\{e^{\theta_i}\}_n$. The parameters $H_{n}^{\boldsymbol{\alpha}}$ are normalisation constants and the factors $e^{\theta_i}+e^{\theta_j}$ in the denominator ensure the correct kinematic pole structure. 

Based on the ansatz (\ref{ansatz}) it is clear that the first step towards solving the form factor equations is the computation of the minimal form factor. For unperturbed IQFTs with diagonal scattering, as considered here, there is a systematic approach for solving the problem, which typically leads to an integral representation of the minimal form factor. 
It is easy to show that, given a solution for the undeformed theory, $F_{\mathrm{min}}(\theta;\boldsymbol{0})$, a consistent solution for the deformed model can be written as
\beq
{F_{\rm{min}}(\theta;{\boldsymbol{\alpha}})}:=\varphi(\theta;\boldsymbol{\alpha}) C(\theta;\boldsymbol{\beta}){F_{\rm{min}}(\theta;\boldsymbol{0})}\,,
\label{key}
\eeq 
with 
\beq 
\varphi(\theta;\boldsymbol{\alpha})  :=\exp\left[{\frac{\theta-i\pi}{2\pi} \sum_{s\in \mathcal{S}} \alpha_s \sinh (s\theta)}\right]\,.
\label{fplus}
\eeq
The formula (\ref{fplus}) for the minimal form factor of $\TTb$-deformed massive integrable theories with diagonal scattering was, to our knowledge, first derived by I. M. Sz\'{e}cs\'{e}nyi in an unpublished work, while its massless version appeared in an investigation in Nambu-Goto theory \cite{Dubovsky:2012wk}, which later turned out to be the $\TTb$ deformation of a free, massless, 2-dimensional scalar field theory.

The function $C(\theta;\bel)$ is a minimal form factor `CDD' factor, that is a function such that any solution can be multiplied by it, while still solving all required equations. It takes the form
\beq C(\theta;\boldsymbol{\beta}):=\exp(\sum_{s\in \mathcal{S}'} \beta_s \cosh(s\theta))\,,\eeq 
where $\mathcal{S}'$ is a subset of the integers. Note that the sum  over $s$ of $ \alpha_s \sinh (s\theta)$ is a sum over the one-particle eigenvalues of local conserved quantities of spin $s$. However, those one-particle eigenvalues are more generically linear combinations of the form $\alpha_s \sinh(s\theta)+ \beta_s \cosh(s\theta)$. Thus the fact that we can always add a sum over $s$ of $\beta_s \cosh (s\theta)$ can be seen as a reflection of this ambiguity in our choice of conserved quantities. It has been recently shown that the structure of the minimal form factor (\ref{key}) is in fact very natural and also found in standard IQFTs such as the sinh-Gordon model \cite{MMF}.

At this stage the usual procedure to obtain higher particle form factors would be to substitute the ansatz (\ref{ansatz}) which solves (\ref{FF1}) and (\ref{FF2}) into equation \eqref{KRE} and find recursive equations for the functions  $Q_n^{\mathcal{O}}(\{\theta_i\}_n;\boldsymbol{\alpha})$ and the constants $H_n^{\boldsymbol{\alpha}}$. We have done this for the Ising field theory and obtained closed solutions for all local fields in \cite{longpaper}. However, also in \cite{longpaper} we observed that the solutions had a factorised structure, as the form factors of the underformed theory, times a new function of the parameters $\bal$ (in \cite{longpaper} we set all $\bel=0$). In the next section, we show that this factorisation is very general and holds beyond the example of the Ising field theory. 

\section{Form Factor Factorisation}
\label{factorisation}
We can report that, for any IQFT with diagonal scattering matrix and a single-particle spectrum, all solutions have the general structure
\begin{equation}
    F_n^{\mathcal{O}}(\theta_1,
\ldots, \theta_n;\boldsymbol{\alpha}) = F_n^{\mathcal{O}}(\theta_1,
\ldots, \theta_n;\bol) \Upsilon_n^{\mathcal{O}}(\theta_1,
\ldots, \theta_n;\boldsymbol{\alpha})\;,
    \label{factor}
\end{equation}
with $\Upsilon_n^{\mathcal{O}}(\theta_1,
\ldots, \theta_n;\boldsymbol{\alpha})$ a non-trivial function of $\boldsymbol{\alpha}$ and $\Upsilon_n^{\mathcal{O}}(\theta_1,
\ldots, \theta_n;\boldsymbol{0})=1$. Here, as we did in \cite{longpaper} for the Ising field theory, we will restrict ourselves to the case $\bel=0$.

The function 
$\Upsilon_n^{\mathcal{O}}(\theta_1,
\ldots, \theta_n;\bal)$ can be computed in complete generality by
plugging (\ref{factor}) into (\ref{ansatz}). 
 This gives rise to the following equations 
\begin{equation}
    H_{n}^{\bal} = H_n^{\bol} \mathcal{H}_{n}^{\bal}\,,\qquad F_{\mathrm{min}}(\theta;\bal) = F_{\mathrm{min}}(\theta;\bol) \varphi(\theta;\bal)\;,
\end{equation}
\begin{equation}
    Q_n^{\mathcal{O}}(\theta_1,\ldots,\theta_n;\boldsymbol{\alpha}) = Q_n^{\mathcal{O}}(\theta_1,\ldots,\theta_n;\bol) \Theta_n^{\mathcal{O}}(\theta_1,\ldots,\theta_n;\boldsymbol{\alpha})\;,
\end{equation}
or equivalently
\begin{equation}
    \Upsilon_n^{\mathcal{O}}(\theta_1,\ldots,\theta_n;\boldsymbol{\alpha}) = \mathcal{H}_{n}^{\bal} \Theta_n^{\mathcal{O}}(\theta_1,\ldots,\theta_n;\boldsymbol{\alpha})\prod_{i<j}^n \varphi(\theta_i - \theta_j;\boldsymbol{\alpha})\;.
    \label{92}
\end{equation}
The quantities $H_n^{\boldsymbol{0}}$, $Q_n^{\mathcal{O}}(\theta_1,
\ldots, \theta_n;\boldsymbol{0})$ and $F_{\mathrm{min}}(\theta;\boldsymbol{0})$ are supposed to satisfy the axioms for the undeformed theory, which means
\begin{equation}
    \mathcal{H}_{n}^{\boldsymbol{0}} = \Theta_n^{\mathcal{O}}(\theta_1,\ldots,\theta_n;\boldsymbol{0}) = \varphi(\theta;\boldsymbol{0}) = 1\;.
\end{equation}
Imposing the form factor equations for the deformed theory we recover the equations\,\footnote{Note that if $\gamma_{\mathcal{O}} = 1$ and $S(\theta) = 1$, the equation for $\Upsilon$ would involve division by zero. Extra care is needed in those cases.}
\begin{equation}
    \varphi(\theta;\boldsymbol{\alpha}) = \Phi_{\boldsymbol{\alpha}}(\theta) \varphi(-\theta;\boldsymbol{\alpha}) = \varphi(2\pi i - \theta;\boldsymbol{\alpha})\;,
\end{equation}
which have already been solved to (\ref{key}), and 
\begin{equation}
    \lim_{\bar{\theta}\rightarrow\theta} \Upsilon_{n+2}^{\mathcal{O}}(\bar{\theta}+\pi i, \theta, \theta_1,\ldots,\theta_n;\boldsymbol{\alpha}) = \frac{1-\gamma_{\mathcal{O}}\prod_{j=1}^n S_{\boldsymbol{\alpha}}(\theta - \theta_j)}{1-\gamma_{\mathcal{O}}\prod_{j=1}^n S_{\bol}(\theta - \theta_j)}\Upsilon_n^{\mathcal{O}}(\theta_1,\ldots,\theta_n;\boldsymbol{\alpha})\;.
\end{equation}
This equation in turn splits into 
\begin{equation}
    \mathcal{H}_{n+2}^{\bal} = \mathcal{H}_{n}^{\bal} \;,
\label{eq:constant_eq2}
\end{equation}
\begin{equation}
    \Theta_{n+2}^{\mathcal{O}}(\theta+i\pi,\theta,\theta_1,\ldots,\theta_n;\boldsymbol{\alpha}) = \mathbb{S}_n\left(\theta - \theta_1, \ldots, \theta - \theta_n ; \gamma_{\mathcal{O}} \right) \Theta_{n}^{\mathcal{O}}(\theta_1,\ldots,\theta_n;\boldsymbol{\alpha})\;,
\label{eq:Theta_eq_gen2}
\end{equation}
where we introduced the quantity
\begin{equation}
     \mathbb{S}_n\left(\theta_1,\ldots, \theta_n ; \gamma_{\mathcal{O}} \right) = \frac{\prod_{j=1}^n S_{\boldsymbol{\alpha}}(\theta_j)^{-1/2} - \gamma_{\mathcal{O}} \prod_{j=1}^n S_{\boldsymbol{\alpha}}(\theta_j)^{1/2}}{\prod_{j=1}^n S_{\bol}(\theta_j)^{-1/2} - \gamma_{\mathcal{O}}\prod_{j=1}^n S_{\bol}(\theta_j)^{1/2}}\;.
\label{eq:doubleS2}
\end{equation}
Notice that as a consequence of the unitarity $S_{\boldsymbol{\alpha}}(-\theta) = S_{\boldsymbol{\alpha}}(\theta)^{-1}$ and crossing symmetry $S_{\boldsymbol{\alpha}}(i\pi-\theta) = S_{\boldsymbol{\alpha}}(\theta)$ of the $S$-matrix, the function \eqref{eq:doubleS2} satisfies the following properties
\begin{equation}
    \mathbb{S}_n\left(\theta_1\pm i\pi,\ldots, \theta_n \pm i\pi ; \gamma_{\mathcal{O}} \right) = \mathbb{S}_n\left(\theta_1 , \ldots , \theta_n ; \gamma_{\mathcal{O}}^{-1} \right)\;,
\label{eq:doubleS_reflection2}
\end{equation}
\begin{equation}
    \mathbb{S}_{n+2}\left(\beta,\beta \pm i\pi,\theta_1 , \ldots , \theta_n ; \gamma_{\mathcal{O}} \right) = \mathbb{S}_n\left(\theta_1 , \ldots , \theta_n ; \gamma_{\mathcal{O}} \right)\;,
\label{eq:doubleS_reduction2}
\end{equation}
which will be useful momentarily.

The recursion \eqref{eq:constant_eq2} for the constant is trivially solved
\begin{equation}
    \mathcal{H}_{n}^{\boldsymbol{\alpha}} = 1\;,
\end{equation}
while the equation \eqref{eq:Theta_eq_gen2} for $\Theta$ requires some more effort, especially for a general undeformed $S$-matrix. Let us give here the solution
\begin{equation}
    \Theta_n^{\mathcal{O}}(\theta_1,\ldots,\theta_n;\boldsymbol{\alpha}) = \sqrt{\prod_{i=1}^n \mathbb{S}_n\left( \theta_i - \theta_1,\ldots , \theta_i - \theta_n ; \gamma_{\mathcal{O}}^{(-1)^i} \right)}\;,
\label{eq:Theta_solution2}
\end{equation}
and check that it indeed solves equation \eqref{eq:Theta_eq_gen2}. The left-hand-side is given by the product of the following three expressions
\begin{equation}
    \sqrt{\mathbb{S}_{n+2}\left(0,i\pi, \theta + i\pi - \theta_1,\ldots , \theta + i\pi - \theta_n ; \gamma_{\mathcal{O}}^{-1} \right)} = \sqrt{\mathbb{S}_n\left(\theta - \theta_1,\ldots , \theta - \theta_n ; \gamma_{\mathcal{O}} \right)}\;,
\end{equation}
\begin{equation}
    \sqrt{\mathbb{S}_{n+2}\left(-i\pi,0 , \theta - \theta_1,\ldots , \theta - \theta_n ; \gamma_{\mathcal{O}} \right)} = \sqrt{\mathbb{S}_n\left(\theta - \theta_1,\ldots , \theta - \theta_n ; \gamma_{\mathcal{O}} \right)}\;,
\end{equation}
and
\begin{equation}
    \sqrt{\prod_{i=1}^n \mathbb{S}_{n+2}\left( \theta_i-\theta-i\pi,\theta_i-\theta,\theta_i - \theta_1,\ldots , \theta_i - \theta_n ; \gamma_{\mathcal{O}}^{(-1)^i} \right)} =  \Theta_n^{\mathcal{O}}(\theta_1,\ldots,\theta_n;\boldsymbol{\alpha})\;,
\end{equation}
where we used the properties \eqref{eq:doubleS_reflection2} and \eqref{eq:doubleS_reduction2}. Taking the product of these three terms indeed yields the right-hand-side of \eqref{eq:Theta_eq_gen2} as long as $\gamma_\mathcal{O}=\gamma_\mathcal{O}^{-1}$ or $\gamma_{\mathcal{O}}=\pm 1$. More explicitly, we can write the final expression
\beqa 
    \Theta_n^{\mathcal{O}}(\{\theta_i\}_n;\boldsymbol{\alpha}) &=& \prod_{i=1}^n \sqrt{\frac{\prod\limits_{j=1}^n S_{\boldsymbol{\alpha}}(\theta_{ij})^{1/2} - \gamma_{\mathcal{O}} \prod\limits_{j=1}^n S_{\boldsymbol{\alpha}}(\theta_{ij})^{-1/2}}{\prod\limits_{j=1}^n S_{\boldsymbol{0}}(\theta_{ij})^{1/2} - \gamma_{\mathcal{O}} \prod\limits_{j=1}^n S_{\boldsymbol{0}}(\theta_{ij} )^{-1/2}}}\nonumber\\
    &=&\prod_{i=1}^n \sqrt{\frac{\sin\left(\frac{1}{2}\sum\limits_{j=1}^n \left[\delta(\theta_{ij})-i\log\Phi_{\boldsymbol{\alpha}}(\theta_{ij})\right] - \frac{\omega_\mathcal{O}}{2} \right)}{\sin\left(\frac{1}{2}\sum\limits_{j=1}^n \delta(\theta_{ij})-\frac{\omega_{\mathcal{O}}}{2}\right)}}\,,
    \label{solution}
\eeqa 
where $S_{\boldsymbol{0}}(\theta)=S_{\boldsymbol{0}}(-\theta)^{-1}=e^{i\delta(\theta)}$ and $\gamma_\mathcal{O}=e^{i\omega_\mathcal{O}}$.  We also assumed that the $S$-matrix is fermionic in the sense that $S_{\boldsymbol{0}}(0)=S_{\boldsymbol{0}}(i\pi)=-1$, so $\delta(0)=\delta(i\pi)=\pi$. As noted above, the solution (\ref{solution}) also requires $\gamma_{\mathcal{O}}=\pm 1$. Thus, strictly speaking, it is valid for local and semi-local fields \footnote{Note that care must be taken for free theories to avoid a zero denominator. This can be done consistently. An example was given in \cite{longpaper}.}. 
\medskip

\subsection{Example}
The factorised structure (\ref{factor}) emerges naturally for free theories. For instance, for the order and disorder fields $(\mu,\sigma)$ in the Ising field theory with $\gamma_\mu=-1$ and $\gamma_\sigma=1$ we have that 
\beqa 
\Theta_n^{\mathcal{O}}(\theta_1,\ldots,\theta_n;\boldsymbol{\alpha})=\sqrt{\prod_{j=1}^n\cos\left(\frac{i}{2}\sum_{i=1}^n \log ( \Phi_{\boldsymbol{\alpha}}(\theta_{ij}))\right)}\,,
\label{free}
\eeqa
with $n$ even for $\mu$ and odd for $\sigma$.
 Note that we have $\Theta_0^{\mathcal{O}}(\boldsymbol{\alpha})=\Theta_1^{\mathcal{O}}(\boldsymbol{\alpha})=1$. The unperturbed form factors in this case are well-known to be proportional to the simple product $\prod_{i<j}\tanh\frac{\theta_{ij}}{2}$ for both fields, with $\mu$ admitting only even particle numbers and $\sigma$ only odd ones. Note that for free theories, $\mathbb{Z}_2$-symmetry is preserved by the generalised $\TTb$ perturbations. In particular, this means that for the Ising field theory, the trace of the stress-energy tensor, usually denoted by $\Theta$ will have non-vanishing even particle form factors, even for particle numbers higher than two. Indeed, these form factors will be also proportional to a factor $\Theta_{2n}^\Theta(\{\theta_{i}\}_{2n};\bal)$ which is identical to (\ref{free}) with cosine replaced by sine.

\vspace{0.1cm}

\section{Correlation Functions and Asymptotics} 
\label{CFA}
One of the most interesting applications of our solutions is to the computation of ground-state two-point functions in the deformed theory. These can be spanned in terms of form factors
For the purposes of this paper, we will just consider the leading contributions to the long and short-distance expansions of the logarithm of a typical (normalised) two-point function $\bra \mathcal{O}(0)\mathcal{O}^\dagger(r)\ket/\bra \mathcal{O}\ket^2$ and will restrict ourselves to fields that only have even-particle form factors. Also for simplicity, we will consider only the case when $\alpha_1:=\alpha\neq 0$ and $\alpha_i=0$ for $i>1$. In this case, the two-particle contribution to the logarithm of the two-point function is
\beqa  
q_2^{\mathcal{O}}(r;\alpha)=
\intop_{-\infty}^\infty \frac{d x}{(2\pi)^2}  \, \left|\Theta_2^{\mathcal{O}}(x;\alpha)\right|^2|\hat{F}_2^{\mathcal{O}}(x;0)|^2 e^{\frac{\alpha x }{\pi}\sinh x} K_0(2r \cosh\frac{x}{2})\,,
\label{cumulant}
\eeqa  
where\,\footnote{Note that for the Ising model $\cos\frac{\delta(\theta)}{2}=0$ in (\ref{17}) but this can be compensated for by appropriate normalisation of the form factors.}
\beq 
\Theta_2^{\mathcal{O}}(\theta;\alpha)=\left\{\begin{array}{cc}
 \left|\frac{\cos\frac{1}{2}\left(\delta(\theta)-\alpha \sinh\theta\right)}{\cos\frac{\delta(\theta)}{2}}\right| & {\rm for} \,\, \gamma_\mathcal{O}=1\\ &\\
   \left|\frac{\sin\frac{1}{2}\left(\delta(\theta)-\alpha \sinh\theta\right)}{\sin\frac{\delta(\theta)}{2}}\right| & {\rm for} \,\, \gamma_\mathcal{O}=-1\\
\end{array}\right. \,,
\label{17}
\eeq 
and $\hat{F}_2^{\mathcal{O}}(x;0)$ is the two-particle form factor normalised by the vacuum expectation value $\bra \mathcal{O}\ket$.
Although this is only the two-particle contribution, it allows us to make some general observations that can then be extended to the full form factor expansion. This is discussed in more detail in \cite{longpaper}.

First, for $\alpha>0$ we see that the exponential term is strongly divergent. This divergence can not be compensated for by the underformed form factor which will generally decay exponentially. For short distances we can interpret this behaviour as indicative of an unusual UV theory. Indeed, it is known \cite{Cardy:2020olv} that for $\alpha>0$ particles acquire a typical (positive) length which makes the UV theory ill-defined. For long distances however, there is an interplay between the length scale, the momentum and the parameter $\alpha$. The integral is divergent above a certain critical value of the rapidity $x$, a cut-off resulting from comparing the dominant exponential decay of the Bessel function and the exponential coming from the minimal form factor
$
e^{-2r\cosh\frac{\Lambda}{2}}=e^{\frac{\alpha \Lambda}{\pi} \sinh\Lambda}\,$. 
For large $\Lambda$, this equation is solved by the principal branch of  Lambert's $W$-function \beq \Lambda=2W_0(\frac{\pi  r}{\alpha})\,.
\eeq 
Thus, for large distances and sufficiently small momentum, the series is still convergent, indicating that the finite length of particles is not visible when probed from a large distance at low momenta/energy. However, if the momentum is high, even with $r$ large the size of particles is again probed and convergence is again spoiled.  

For $\alpha<0$ on the other hand there is rapid convergence both for short and long distances. One possible interpretation of this behaviour is that for $\alpha<0$, particles acquire a (negative) length. This may be understood as follows: the description in terms of positive/negative lengths comes from the relationship with the hard rod model \cite{hard}. The hard rod model is a classical model describing the interaction amongst hard rods of length $\ell$. The sign of $\ell$ determines the scattering shift associated with the collision of two rods, as well as the thermodynamic properties of the system. In \cite{hard} it was shown that the non-relativistic limit of the TBA/GHD equations is described by the hard rod model. This viewpoint leads to another interpretation of the $\TTb$-deformation, namely that its non-relativistic version corresponds to a change in particle (rod) length. Since particle (rod) length is only a model parameter related to the scattering phase, it can be chosen to be either positive or negative, hence our interpretation of $\alpha>0$ as positive length and $\alpha<0$ as negative length. 

In the IR due to the presence of a mass scale, the theory flows to a trivial IR point, just as in the unperturbed case. In the UV, the form factor expansion needs to be resummed in order to obtain the exact leading behaviour. However, since $K_0(2r \cosh\frac{\theta}{2})\approx -\log r$ for $r\ll1$ and a similar behaviour is found for every contribution in the form factor series, the $n$-particle contribution  will scale as $-z_{n}^\mathcal{O}(\alpha) \log r$ with coefficient $z_{n}^\mathcal{O}(\alpha) $ given in terms of a (multiple) integral in the rapidities. For instance, for two particles, the contribution scales as 
\beq 
q_2^{\mathcal{O}}(r\ll 1;\alpha<0)=-z_{2}^\mathcal{O}(\alpha) \log r \,,
\eeq 
with 
\beq
 \! \!\! \! z_{2}^\mathcal{O}(\alpha) = \intop_{-\infty}^\infty \frac{d\theta}{(2\pi)^2} \, \left|\Theta_2^\mathcal{O}(\theta;\alpha)\right|^2 |\hat{F}_2^{\mathcal{O}}(\theta;0)|^2 e^{\frac{\alpha \theta}{\pi}\sinh\theta}\,.
\eeq 
The rapidly decaying $\alpha$-dependent exponential has the effect of producing coefficients $z_{n}^\mathcal{O}(\alpha)$ such that $
z_{n}^\mathcal{O}(\alpha)  \leq z_{n}^\mathcal{O}(0)$ for  $\alpha<0$,
whenever $z_{n}^\mathcal{O}(0) $ is well-defined\,\footnote{It is worth noting that this formula only makes sense for $\alpha=0$ if the unperturbed form factor decays exponentially for large $\theta$. This is not the case for the fields $\mu, \Theta$ in the Ising field theory, for instance.}. 
In other words, the UV behaviour is characterised by a power law just as happens in CFT, but this power is $\alpha$-dependent and tends to zero as $\alpha\rightarrow -\infty$.


\section{$c$-Theorem and Numerics} 
\label{cTheorem}
Another avenue that can be explored is Zamolodchikov's $c$-theorem, which generally relates the connected part of the two-point function of the trace of the stress-energy tensor\,\footnote{Note that the field $\Theta$ -- the trace of the stress-energy tensor -- should not be confused with the function $\Theta_n^\mathcal{O}(\theta;\alpha)$ introduced earlier.} with the change in the central charge between the IR and UV fixed points
\begin{equation}
    c(\alpha):=c^{UV} - c^{IR}=  \frac{3}{2} \intop_0^{\infty} dr \, r^3 \langle \Theta(0) \Theta(r) \rangle_c\,.
\label{eq:c_theorem}
\end{equation}
For $\alpha = 0$, this equation gives the value of $c^{UV}$, since $c^{IR} = 0$, in agreement with the fact that a massive theory flows in the IR to a trivial gapped fixed point. A $\TTb$ perturbation alters the UV behaviour of the the theory. For $\alpha < 0$ we still expect to have $c^{IR} = 0$ and so equation \eqref{eq:c_theorem} will give some finite value which can be interpreted as the deformation of $c^{UV}$ under the $\TTb$ perturbation.
\begin{figure}[h!]
\begin{center}
	\includegraphics[width=7cm]{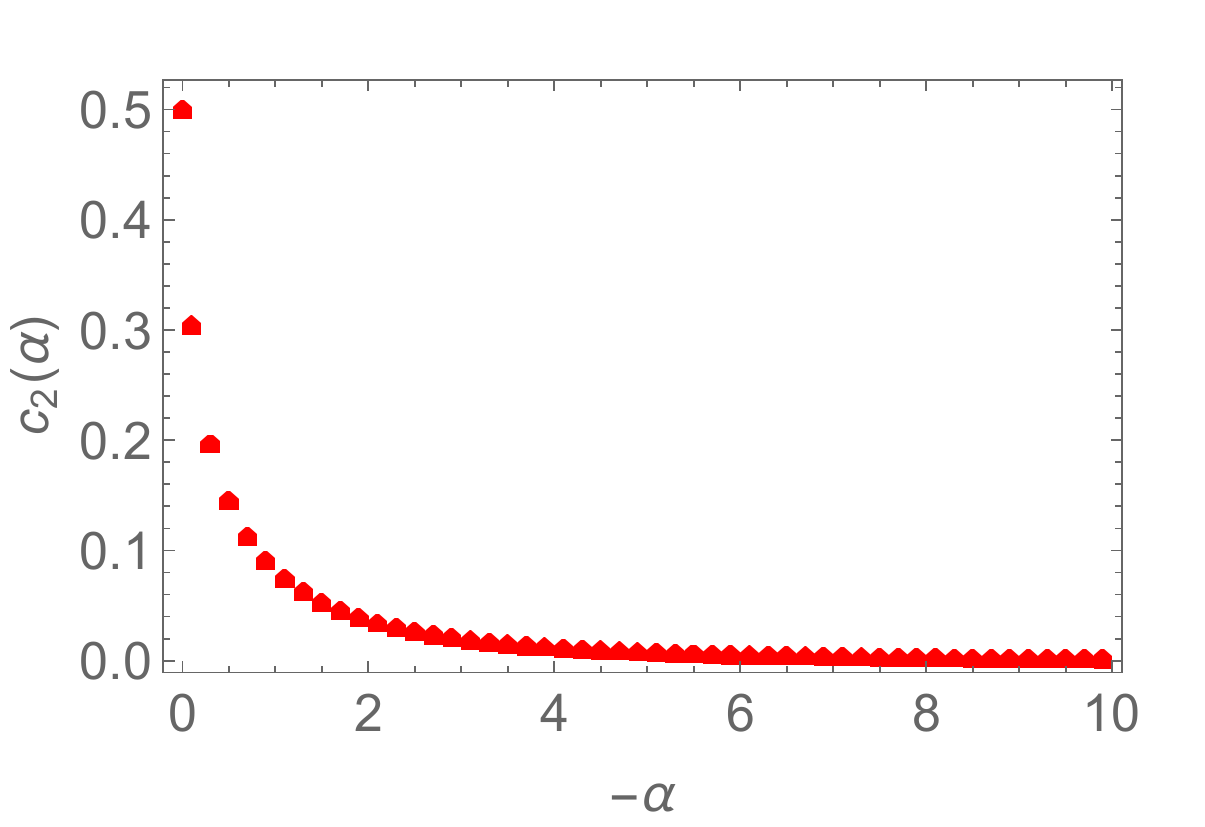}
				    \caption{The functions $c_2(\alpha)$ for the Ising model.}
				     \label{fig1}
    \end{center}
    \end{figure}
    This quantity will now depend on $\alpha$ and behave as shown in Fig.~\ref{fig1}. The trace of the stress-energy tensor is a local field hence its two-particle form factor does not have a kinematic pole. As we have shown in \cite{longpaper}, for the Ising model and a $\TTb$ perturbation, it takes the form
\beq 
F_2^\Theta(\theta;\alpha):=2\pi \left|\frac{\sin\left(\frac{\alpha}{2}\sinh\theta\right)}{\frac{\alpha}{2}\sinh\theta} \right|  \frac{F_{\bf min}(\theta; \alpha)}{F_{\bf min}(i\pi;\alpha)}\,.
\eeq 
where the oscillatory factor is of the same type found earlier.  In the two-particle approximation, after integration in $r$ and in one rapidity variable, the integral (\ref{eq:c_theorem}) gives
\begin{equation}
 c_2(\alpha) = \frac{3}{8} 
     \intop_{ - \infty}^{+ \infty} dx \left[\frac{\sin\left(\frac{\alpha}{2}\sinh x\right)}{{\alpha}\cosh^3\frac{x}{2}} \right]^2 e^{\frac{\alpha}{\pi}x \sinh{x}}.
\end{equation}
The resulting function is evaluated numerically in Fig.~1 where we see that it flows from the unperturbed value $1/2$ to zero for $-\alpha$ large. The oscillatory dependence is entirely washed out by the rapidly decaying exponential. 
\vspace{0.05cm}
    
\section{Conclusions, Discussion and Outlook}
\label{conclu}
In this paper we have proposed a form factor program for IQFTs perturbed by $\TTb$ and descendants thereof. We have shown that the form factor equations can be solved exactly for a large class of theories and fields, and that these solutions span correlation functions in the usual way. We find that, compared to the unperturbed theory, all form factor solutions are ``dressed" by oscillatory functions of the rapidities and perturbation parameters. 

Our derivation is subject a number of constraints and choices that we would like to now discuss in some detail. As we have seen, we requires the factors $\gamma_{\mathcal{O}}=\pm 1$ and have chosen $\bel=0$. In addition, in order for (\ref{eq:doubleS2}) to hold, we have assumed all spins $s$ to be odd. This assumption is rather natural since the local conserved charges of most IQFTs are indeed odd. In that case, the following condition holds,
$$\varphi(\theta;\bal)\varphi(\theta+i\pi;\bal)=\Phi_{\bal}^{1/2}(\theta)\,.$$
This constraint can however be easily lifted by modifying the definition of the function  (\ref{eq:doubleS2}). 

An additional implicit assumption, built into the ansatz (\ref{ansatz}) is that all form factors have kinematic poles. However, local fields, such as the trace of the stress-energy tensor, do not have a kinematic pole for their two-particle form factor but have kinematic poles for higher particle numbers. 
We may then ask what changes in this derivation for local fields. In fact, for interacting theories, the derivation presented here goes through with $\gamma_{\mathcal{O}}=1$ and for particle numbers $n>2$,  since the two-particle and higher particle solutions have slightly different structures.  In fact, in order to make the solution $\Theta_n^{\mathcal{O}}(\theta_1,\ldots,\theta_n;\bal)$ found here consistent with a two-particle form factor without kinematic pole and with the kinematic residue equations we will typically need to multiply our solution (\ref{solution}) by an additional function $g_n(\theta_1,\ldots,\theta_n;\bal)$ with the property that it solves all form factor equations trivially, that is
\beq 
g_{n+2}(\theta+i\pi,\theta,\theta_1,\ldots,\theta_n;\bal)=g_{n}(\theta_1,\ldots,\theta_n;\bal)\,,
\label{gen2}
\eeq 
while being symmetric and $2\pi i$-periodic in all variables. This function will be specific to each theory. For the Ising field theory we have discussed the construction of this function in \cite{longpaper}.

An unusual feature of our form factor solutions is that the function (\ref{solution}) which acts as a kind of ``dressing" of the unperturbed form factors, involves a square root. We think that this may be related to the fact that a $\TTb$ perturbation can be interpreted in the TBA sense as a state-dependent change of the inverse temperature (or generalised temperatures for more general perturbations). {It seems therefore natural} that a relationship between finite temperature form factors and $\TTb$-deformed form factors should exist. Interestingly, form factors at finite temperature {also} involve square roots, as can be seen for instance for the Ising model in \cite{Doyon:2005jf,Doyon:2006pv}. Closely related are finite volume form factors, which are also typically dressed by square root factors \cite{Pozsgay:2007gx,Pozsgay:2007kn}. Both for finite temperature and for finite volume form factors, the square root factors are functions of TBA quantities. It would be very interesting to investigate this connection further. This viewpoint also gives a new interpretation to the $c$-function presented in Fig.\ref{fig1}, which has a very similar form as the TBA scaling function, if we replace inverse temperature by $-\alpha$. 

The asymptotics of correlation functions is consistent with the picture that point-like degrees of freedom (particles) in unperturbed theories become objects of finite (positive, for $\alpha>0$, negative for $\alpha<0$) length after the perturbation. In this light, it is clear the UV theory is not a standard QFT whereas in the presence of a mass scale, the IR theory is as usual trivial. Correlation functions obtained from form factors reflect these features by diverging at short distances with $\alpha>0$ and converging very rapidly for $\alpha<0$, while displaying power-law scaling with $\alpha$-dependent exponents. 

 Together with \cite{longpaper,entropyTTb,MMF}, this work provides a new viewpoint on generalised $\TTb$-perturbed theories as well as new computational tools. Many different problems can now be addressed, from the study of correlation functions in interacting theories, to that of entanglement measures. There remain also many interesting open questions, e.g. what is the interpretation of the function $c(\alpha)$ and of the powers $z^{\mathcal{O}}(\alpha)$ characterising the short-distance scaling of correlators? We hope to return to these problems in the near future.
\medskip

\noindent {\bf Acknowledgments:} The authors thank John Donahue, Benjamin Doyon, Fedor Smirnov, Roberto Tateo and Alexander Zamolodchikov for useful discussions.. Olalla A. Castro-Alvaredo thanks EPSRC for financial support under Small Grant EP/W007045/1. The work of Stefano Negro is partially supported by the NSF grant PHY-2210349 and by the Simons Collaboration on Confinement and QCD Strings. Fabio Sailis is grateful for his PhD Studentship which is funded by City, University of London. This project was partly inspired by a meeting at the Kavli Institute for Theoretical Physics (Santa Barbara) in September 2022. Olalla A. Castro-Alvaredo and Stefano Negro thank the Institute for financial support from the National Science Foundation under Grant No. NSF PHY-1748958, and hospitality during the conference ``Talking Integrability: Spins, Fields and  Strings", August 29-September 1 (2022) and the related extended program on ``Integrability in String, Field and Condensed Matter Theory", August 22-October 14 (2022).

\bibliography{Ref}

\end{document}